\documentstyle[epsf]{l-aa}
\begin{document}

\thesaurus{02.04.2; 06.01.1; 06.05.1; 06.09.1; 06.18.2}
\title{
New Solar Models Including Helioseismological Constraints and 
 Light-Element Depletion}

\author{O. Richard\inst{1} \and S. Vauclair\inst{1} \and 
 C. Charbonnel\inst{1} \and W.A. Dziembowski\inst{2}}

\institute{
 Observatoire Midi-Pyr\'en\'ees
   14, avenue Edouard Belin, 31400 Toulouse, France \and
 Copernicus Astronomical Center, Warsaw, Poland
}

\markboth{}{O. Richard et al.}

\offprints{S. Vauclair}

\date{Received ; accepted }
\maketitle
\begin{abstract}
We have computed new solar models using the same stellar evolution code as
described in Charbonnel, Vauclair and Zahn (1992). This code, originating
from Geneva, now includes the computation of element segregation for helium
and 12 heavier isotopes. It may also include any type of mixing of the
stellar gas, provided this mixing can be parametrized  with an effective
diffusion coefficient as a function of radius. Here we introduced
rotation-induced mixing as prescribed by Zahn (1992).
We present five solar models: 1) the standard model, computed with heavy
element
abundances as given by Grevesse (1991); 2) a model including pure element
segregation (no mixing outside the convective zone) with 
Grevesse (1991)
as initial abundances; 3) same model as (2), but iterated so that the final
abundances are those of Grevesse (1991); 4) a model with both element
segregation and rotation-induced mixing, leading to lithium and beryllium
depletion consistent with the observations, with Grevesse (1991) as initial
abundances; 5) same model as (4) but iterated to obtain Grevesse (1991) as
final abundances. This model (5) now represents our best new solar model
consistent with the observations.

The $u =  {P \over \rho  }$ function computed as a function of radius in these new solar
models are compared to the helioseismological results obtained for the same
function by Dziembowski et al (1994). Improving the physics of the models
leads to a better consistency with helioseismology. In our best model (5),
which includes both segregation and mixing, the relative difference in the
$ u $ function between the model and the helioseismological results is smaller
than 0.5 per cent at all radii except at the center and the surface.
Meanwhile lithium is depleted by a factor 155 and beryllium by a factor
2.9, which is consistent with the observations. The bottom of the
convective zone lies at a fractional radius of 0.716, consistent with
helioseismology. The neutrino fluxes are not decreased in any of these
models.

The models 
including the computations of
element segregation 
lead to a present surface helium abundance of:
$ Y_{surf} $ between 0.248 and 0.258, which is in  
satisfactory agreement with the value derived from 
helioseismology.

\keywords{Physical data and processes : diffusion --
Sun : abundances - evolution - interior - 
rotation}

\end{abstract}
\section{Introduction}
The Sun is by far the most well known of all the stars. Its mass, radius,
luminosity and age have been determined with a high degree of precision
(Table 1). The mass is obtained from the motion of planets, the radius from
eclipses (the value given in table 1 is reduced to an optical depth
$ \tau  $=2/3), the luminosity from measurements of the solar constant above the
earth's atmosphere. Some discussion remains about the solar age: it is
generally taken as 4.6 billion years although Guenther 
(1989) and Demarque and Guenther (1991) suggest a
smaller age consistent with the oldest meteorites.
\begin{table}[h]
\caption{Solar parameters}
\begin{tabular}{lc}
\hline\noalign{\smallskip}
 mass & $(1.9891 \pm 0.0004) \times  10^{33} $g \\ 
 radius & $ (6.959 \pm 0.001) \times  10^{10} $ cm \\ 
 luminosity & $ (3.851 \pm 0.005) \times  10^{33} $ ergs.s$^{-1}$ 
\\ 
 age & $ 4.6 \pm 0.15 $ Gyr \\ 
 mass loss & $ 2 \times  10^{-14}$ M$_{\odot}$.yr$^{-1}$\\ 
\noalign{\smallskip}
\hline
\end{tabular}
\end{table}
The photospheric solar element abundances are now precisely known, after
the studies by Anders and Grevesse (1989) modified by Grevesse (1991). 
See also Grevesse and Anders (1991).
Lithium and beryllium are both depleted in the Sun with ratios :
  $$  
Li/Li_{o}  = 1 / 140 \quad ; \quad Be/Be_{o}  =  1/2 
  $$ 
with an relative uncertainty of 
30\% in both cases.
Any consistent solar model must account for these depletion factors.

The study of the internal structure of the Sun entered a new age with the
birth of helioseismology. Millions of solar p-modes have been detected 
(including the ($2l+1$) multiplets). 
An inversion of the measured frequencies yields accurate and 
detailed information about such structural functions as 
pressure, $ p \ (r) $, and density,
$ \rho  \ (r) $, in the Sun's interior. A particularly high 
precision is achieved in the determination of $ u =  {p \over  \rho } 
$\, throughout the Sun, and in the 
localization of the bottom of the convective envelope. No 
assumption regarding the transport of energy and chemical 
elements is introduced at this stage of seismic sounding. 
The only essential assumption is that of mechanical 
equilibrium, which is partially testable by means of 
helioseismology.

Christensen-Dalsgaard et al. (1993) have shown that 
gravitational settling and element mixing processes 
significantly affect the calculated speed of sound near the 
bottom of the convective envelope. These processes lead to a 
lower surface helium abundance - a quantity which may also be 
directly inferred form helioseismic data if the equation of 
state is specified. The results of many independent 
inversions clearly demonstrate that helium settling must take 
place in the Sun. A question we ask in this paper is whether 
helioseismology provides useful constraints on the mixing 
processes.

An important result of  helioseismology is the precise 
determination of the bottom of the solar convective zone:
$  {r_{cz} \over R_{\odot} }~=~0.713~\pm~0.003 $  (Christensen-
Daalsgard et al. 1993). This value corresponds to that 
obtained from the Schwarzschild criterium, leading to a 
strong constraint on overshooting (section~3). In the following we 
show that, 
contrary to a common idea generally spread among solar physicists, this
constraint on overshooting is not a problem for the explanation of the
lithium depletion in the Sun. 
In any case, explaining
 the lithium depletion in the Sun by overshooting would not 
 be consistent with 
the lithium observations in other stars 
 (e.g. galactic clusters).

Charbonnel, Vauclair and Zahn (1992)
(CVZ) and Charbonnel et al. (1994)
 showed that the lithium deficiency in solar type stars can be
accounted for by rotation-induced mixing. We will show here that such a
mixing can also account for Li and Be depletion in the Sun, 
without destroying the consistency with helioseismology.

\begin{table*}
\caption{Initial parameter and main physical surface parameter of solar
models.}
\begin{flushleft}
\begin{tabular}{ccccccccc}
\hline\noalign{\smallskip}
 \multicolumn{1}{c}{}
& \multicolumn{1}{c}{Y$_0$}
& \multicolumn{1}{c}{$\alpha$}
& \multicolumn{1}{c}{$Y_{surface}$}
& \multicolumn{1}{c}{$X_{surface}$}
& \multicolumn{1}{c}{L}
& \multicolumn{1}{c}{R}
& \multicolumn{1}{c}{Li/Li$_0$}
& \multicolumn{1}{c}{Be/Be$_0$}
\\
 & & & & & (10$^{34}$erg.s$^{-1}$) & (10$^{11}$cm) & &  \\
\noalign{\smallskip}
\hline\noalign{\smallskip}
Model 1 & 0.2782 & 1.652 & 0.2782 & 0.7028 & 0.385145 & 
0.695976 & 1 & 1 \\
Model 2 & 0.2762 & 1.776 & 0.2477 & 0.7341 & 0.385154 &
0.696368 & 1/2.89 & 1/1.17 \\
Model 3 & 0.2798 & 1.789 & 0.2513 & 0.7297 & 0.385143 &
0.695982 & 1/3.50 & 1/1.17 \\
Model 4 & 0.2770 & 1.761 & 0.2563 & 0.7252 & 0.385131 &
0.695980 & 1/124.58 & 1/2.88 \\
Model 5 & 0.2793 & 1.768 & 0.2584 & 0.7226 & 0.384993 &
0.695849 & 1/155.03 & 1/2.91 \\
\noalign{\smallskip}
\hline
\end{tabular}
\end{flushleft}
\end{table*}
\begin{table*}
\caption{Main physical parameter of solar models at the base
of the convective zone and at the center.} 
\begin{flushleft}
\begin{tabular}{ccccccccc}
\hline\noalign{\smallskip}
 \multicolumn{1}{c}{}
& \multicolumn{1}{c}{${r_{cz} \over {R_{\odot}}}$}
& \multicolumn{1}{c}{T$_{cz}$}
& \multicolumn{1}{c}{$\rho_{cz}$}
& \multicolumn{1}{c}{Y$_c$}
& \multicolumn{1}{c}{X$_c$}
& \multicolumn{1}{c}{T$_c$}
& \multicolumn{1}{c}{$\rho_c$}
& \multicolumn{1}{c}{P$_c$}\\
 & & (10$^6$K) & (g.cm$^{-3}$) & & &(10$^6$K) & (g.cm$^{-3}$) & ( 
dyn.cm $^{-2
}$ ) \\ 
\noalign{\smallskip}
\hline\noalign{\smallskip}
Model 1 & 0.725 & 2.100 & 0.166 & 0.6346 & 0.3459 & 15.56 & 150.66
& 2.303 $10^{17}$ \\
Model 2 & 0.716 & 2.158 & 0.185 & 0.6416 & 0.3383 & 15.63 & 153.81
& 2.344 $10^{17}$ \\
Model 3 & 0.714 & 2.178 & 0.189 & 0.6464 & 0.3326 & 15.70 & 154.17
& 2.345 $10^{17}$ \\
Model 4 & 0.717 & 2.162 & 0.185 & 0.6431 & 0.3368 & 15.63 & 154.17
& 2.350 $10^{17}$ \\
Model 5 & 0.716 & 2.175 & 0.188 & 0.6465 & 0.3328 & 15.67 & 154.53
& 2.350 $10^{17}$ \\
\noalign{\smallskip}
\hline
\end{tabular}
\end{flushleft}
\end{table*}
The recipe to construct solar models is well known and has been explained
many times in the literature
(Cox, Guzik and Kidman 1989, Bahcall and Pinsonneault 1992, 
Turck-Chieze and Lopes 1993, Proffitt 1994 etc.).
We begin with a homogeneous 1M$ _{\odot} $ model on
the zero-age main sequence. Then the model is evolved step by step,
by taking into account
the modification of the chemical composition due to nuclear reactions. At
the age of the Sun, it must reproduce the data given in table 1 within the
observed uncertainties. To obtain such a high precision, two parameters are
traditionally adjusted, with an iteration prodedure: the 
${}^4$He mass
fraction Y and the ratio of the convective mixing length to the pressure
scale height, generally referred to as $ \alpha  $.

The computation of the standard solar models includes the assumption that
the stellar gas as a whole is in hydrostatic equilibrium. This fundamental
``first equation'' of the
internal structure of stars assumes that a blob of stellar gas is in
equilibrium due to the effect of gravity downwards and pressure gradient
upwards, which is correct in first approximation. It does not take however
into account the fact that the stellar gas is
a mixture of many different species, which do not have the same
weight. This is the problem of element segregation, which indeed is a
fundamental process inherent to the stellar structure.

In this paper we present new 
solar models computed with a stellar
evolution code including element segregation and mixing, as described in
CVZ. The physics included in these computations is discussed in section 2,
a discussion about helioseismology appears in section 3 and  the
results are given  in section 4.

Five solar models will be discussed: 1) the best standard model 2) a model
including pure element segregation (no mixing outside the convective zone)
with  initial abundances corresponding to the Grevesse (1991)
mixture; 3) same model as (2), but
iterated so that the final abundances are those of Grevesse (1991); 4) a
model with both element segregation and rotation-induced mixing, leading to
lithium and beryllium depletion consistent with the observations, with
Grevesse (1991) as initial abundances; 5) same model as (4) but iterated to
obtain Grevesse (1991) as final abundances. This model (5) now represents our
best new solar model consistent with the observations.

All these models are compared to the helioseismological results obtained by
Dziembowski et al (1994). It is very encouraging to see that improving the
physics of the models leads to a very good consistency with helioseismology.
In our best model (5), which includes both segregation and mixing, the
relative difference in the u function between the model and the
helioseismological results is smaller than 0.5 per cent at all radii except
at the center and the surface. Meanwhile lithium is depleted by a factor
150 and beryllium by a factor 2.9, which is consistent with the
observations. The bottom of the convective zone lies at a fractional radius
of 0.716, consistent with helioseismology.

The computed neutrino fluxes will be presented in the tables of results for
each models. They are not decreased in any of them. These results will be
shortly discussed but a complete discussion of the solar neutrino problem
is out of the scope of the present paper.

\section{The computations}

\begin{figure}
\begin{center}
\epsfxsize=9.5cm
\epsfysize=18.5cm
\epsfbox{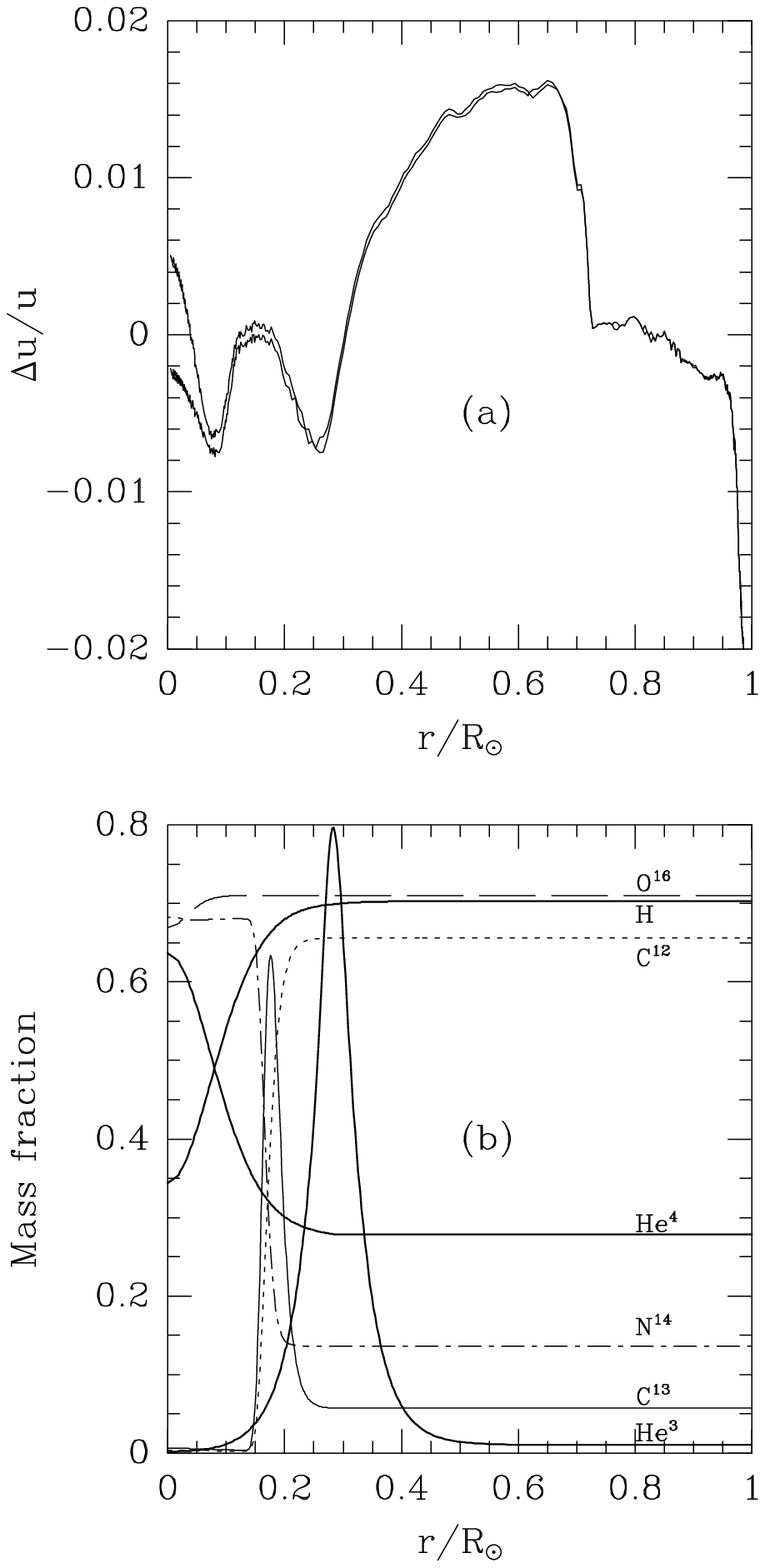}
\end{center}
\caption{
Standard Solar model (model~1). The top graph represents
the difference between the
$ u $ function $ (u =  {P \over \rho  } )$ deduced from
helioseismology and the computed one. The ordinates
represents : $  {\Delta u \over  u} =  {u \ \hbox{(seismic)} - u \
\hbox{(model)}
\over u \ \hbox{(seismic)} }$\hfil\break
The bottom graph shows the abundance profiles for
some elements in the two models (The mass fractions are
multiplied by 250 for $ ^{3} $He, by 200 for $ ^{12} $C, by
1000 for $ ^{13} $C, by 140 for $ ^{14} $N and by 75 for $ ^{16}
$O).}
\end{figure}

\subsection{Input microphysics}

-- Equation of state :
The present solar models are computed with the equation of state
developped by Hummer \& Mihalas (1988), Mihalas et al. (1988), D\"appen
et al. (1988), hereafter MHD. 
The MHD equation of state is based on the free-energy minimization
method which implies that it is thermodynamically consistent.
It treats pressure ionization carefully and takes into account non ideal 
effects such as Coulomb correction to pressure,
pressure  due  to  partially  degenerate  electrons 
and correction for size of particles. 
It also  includes a large number of atomic, ionic and molecular
species,  with  detailed  partition  functions, containing weighted
occupation probabilities.   
As shown by Christensen-Dalsgaard et al. (1988),  
the MHD equation of state highly 
improves the solar models in the helioseismological context.

-- Opacities : 
We use the OPAL radiative opacities by Iglesias et al. (1992) which
include the spin-orbit interactions for Fe and relative metal abundances 
based on Grevesse (1991). These tables are complemented at low temperatures  
below  10\,000K  with the atomic and molecular opacities by Kurucz (1991). 

-- Nuclear reactions :
For hydrogen-burning we consider the tree pp chains and the CNO
tri-cycle. We use the thermonuclear reaction rates Caughlan \& Fowler
(1988). 
Screening factors for the reaction rates are taken into account according 
to the analytical prescription by Graboske et al. (1973).

The key reaction for the generation of high energy neutrinos and for the
theoretical neutrino flux to be compared to the chlorine experiments
results is $^7$Be(p,$\gamma )^8$B. 
Caughlan \& Fowler (1988) give a value of 0.0240 keV-barns for the 
corresponding low energy cross section factor, S$_{17}$(0), which is 
extrapolated from experimentally measured reaction cross sections.

\subsection{Element Segregation}

The process of element segregation in stars (also referred to as
``microscopic diffusion'')
represents a basic physical process inherent to the stellar structure. As
soon as the stars form out of gas clouds, they built density, pressure
and temperature gradients throughout. Under such conditions, the various
chemical species present in the stellar gas move with respect to one
another, unless macroscopic motions force the chemical homogeneization.

Although recognized by the pioneers of the study of stellar structure
(Eddington, 1916 and 1926, Chapman, 1917), this process was long
forgotten in the computations of stellar models, except
for white dwarfs (Schatzman, 1945).
Only with the discovery of large abundance anomalies in main-sequence
type stars (the so-called Ap and Am stars), which present characteristic
variations of chemical elements 
with the effective temperature, was microscopic diffusion
brought into light fifty years later (Michaud, 1970, see other references
in Vauclair and Vauclair, 1982).

At that time, the effects of microscopic diffusion were supposed to be
important only when the diffusion time scale was smaller than the
stellar age. In the Vauclair and Vauclair (1982) review paper, Fig. 1
shows the regions in the HR diagram where microscopic diffusion could
lead to ``observable'' abundance variations. The Sun was excluded,
although at the border of the ``permitted domain".
In the present days, due to helioseismology, abundance variations
of the order of a few percent become indirectly
detectable: we have entered a new area
in this respect.

Several authors have computed the gravitational and thermal diffusion of
helium and heavier elements in the Sun with various approximations
(see references in Michaud and Vauclair, 1991). More recently the
influence of diffusion on the solar oscillation modes and on the solar
neutrino fluxes have been studied in various ways (Cox, Guzik, Kidman, 1989;
Bahcall and Pinsonneault, 1992; Proffitt, 1994; Thoul, Bahcall and
Loeb, 1994).
\begin{table*}
\caption{Initial abondances, final surface abondances, and Grevesse
mixture in mass fraction.}
\begin{flushleft}
\begin{tabular}{lcccccccc}
\hline\noalign{\smallskip}
 \multicolumn{1}{c}{}
& \multicolumn{4}{c}{Model 2}
& \multicolumn{4}{c}{Model 3}\\
\noalign{\smallskip}
\cline{2-9}\noalign{\smallskip}
 & Initial & Grevesse & Final & Grevesse &
Initial & Grevesse & Final & Grevesse \\
 & mixture & (1991) & mixture & (1991) & mixture & (1991) & mixture &
(1991) \\
 & used & with same & obtained & with same & used & with same & obtained & with
same \\
 & (I2) & Y than I2 & (F2) & Y than F2 & (I3) & Y than I3 & (F3) & Y
than F3 \\
\noalign{\smallskip}
\hline\noalign{\smallskip}
X & 0.704800 & 0.704800 & 0.734103 & 0.733309 & 0.700428 & 0.701185 &
0.729746 & 0.729653 \\
X$_{He^{3}}$ & 0.000042 & 0.000042 & 0.000040 & 0.000044 & 0.000046 & 0.000042 &
0.000043 & 0.000044 \\
Y & 0.276200 & 0.276200 & 0.247691 & 0.247691 & 0.279815 & 0.279815 &
0.251347 & 0.251347 \\
X$_{C^{12}}$ & 0.003288 & 0.003288 & 0.003084 & 0.003421 & 0.003593 & 0.003271 & 0.003373 & 0.003404 \\
X$_{C^{13}}$ & 0.000057 & 0.000057 & 0.000053 & 0.000059 & 0.000063 & 0.000057 &
0.000059 & 0.000059 \\
X$_{N^{14}}$ & 0.000976 & 0.000976 & 0.000919 & 0.001015 & 0.001062 & 0.000971 &
0.001001 & 0.001010 \\
X$_{N^{15}}$ & 0.000004 & 0.000004 & 0.000004 & 0.000004 & 0.000004 & 0.000004 &
0.000004 & 0.000004 \\
X$_{O^{16}}$ & 0.009494 & 0.009494 & 0.008971 & 0.009878 & 0.010313 & 0.009445 &
0.009754 & 0.009828 \\
X$_{O^{17}}$ & 0.000007 & 0.000007 & 0.000007 & 0.000007 & 0.000007 & 0.000007 &
0.000007 & 0.000007 \\
X$_{O^{18}}$ & 0.000023 & 0.000023 & 0.000022 & 0.000024 & 0.000025 & 0.000023 &
0.000023 & 0.000024 \\
X$_{Ne^{20}}$ & 0.001603 & 0.001603 & 0.001603 & 0.001668 & 0.001657 & 0.001595
& 0.001657 & 0.001660 \\
X$_{Ne^{22}}$ & 0.000140 & 0.000140 & 0.000140 & 0.000146 & 0.000145 & 0.000140
& 0.000145 & 0.000145 \\
X$_{Mg^{24}}$ & 0.000497 & 0.000497 & 0.000497 & 0.000517 & 0.000514 & 0.000494
& 0.000514 & 0.000515 \\
X$_{Mg^{25}}$ & 0.000068 & 0.000068 & 0.000068 & 0.000071 & 0.000071 & 0.000068
& 0.000071 & 0.000071 \\
X$_{Mg^{26}}$ & 0.000081 & 0.000081 & 0.000081 & 0.000085 & 0.000084 & 0.000081
& 0.000084 & 0.000084 \\
Z/X & 0.026958 & 0.026958 & 0.024800 & 0.025910 & 0.028207 & 0.027097 &
0.025909 & 0.026040 \\
\noalign{\smallskip}
\hline
\end{tabular}
\end{flushleft}
\begin{flushleft}
\begin{tabular}{lcccccccc}
\hline\noalign{\smallskip}
 \multicolumn{1}{c}{}
& \multicolumn{4}{c}{Model 4}
& \multicolumn{4}{c}{Model 5}\\
\noalign{\smallskip}
\cline{2-9}\noalign{\smallskip}
 & Initial & Grevesse & Final & Grevesse &
Initial & Grevesse & Final & Grevesse \\
 & mixture & (1991) & mixture & (1991) & mixture & (1991) & mixture &
(1991) \\
 & used & with same & obtained & with same & used & with same & obtained & with
same \\
 & (I2) & Y than I2 & (F2) & Y than F2 & (I3) & Y than I3 & (F3) & Y
than F3 \\
\noalign{\smallskip}
\hline\noalign{\smallskip}
X & 0.704016 & 0.704016 & 0.725216 & 0.724734 & 0.701229 & 0.701714 &
0.722576 & 0.722551 \\
X$_{He^{3}}$ & 0.000042 & 0.000042 & 0.000076 & 0.000043 & 0.000018 & 0.000042 &
0.000054 & 0.000043 \\
Y & 0.276984 & 0.276984 & 0.256266 & 0.256266 & 0.279286 & 0.279286 &
0.258449 & 0.258449 \\
X$_{C^{12}}$ & 0.003284 & 0.003284 & 0.003149 & 0.003381 & 0.003498 & 0.003273 & 0.003355 & 0.003371 \\
X$_{C^{13}}$ & 0.000057 & 0.000057 & 0.000054 & 0.000059 & 0.000062 & 0.000057 &
0.000059 & 0.000059 \\
X$_{N^{14}}$ & 0.000975 & 0.000975 & 0.000938 & 0.001003 & 0.001035 & 0.000971 &
0.000996 & 0.001000 \\
X$_{N^{15}}$ & 0.000004 & 0.000004 & 0.000004 & 0.000004 & 0.000004 & 0.000004 &
0.000004 & 0.000004 \\
X$_{O^{16}}$ & 0.009483 & 0.009483 & 0.009142 & 0.009762 & 0.010052 & 0.009452 &
0.009693 & 0.009733 \\
X$_{O^{17}}$ & 0.000007 & 0.000007 & 0.000007 & 0.000007 & 0.000007 & 0.000007 &
0.000007 & 0.000007 \\
X$_{O^{18}}$ & 0.000023 & 0.000023 & 0.000022 & 0.000024 & 0.000025 & 0.000023 &
0.000024 & 0.000024 \\
X$_{Ne^{20}}$ & 0.001602 & 0.001602 & 0.001602 & 0.001649 & 0.001641 & 0.001596
& 0.001641 & 0.001644 \\
X$_{Ne^{22}}$ & 0.000140 & 0.000140 & 0.000140 & 0.000144 & 0.000144 & 0.000140
& 0.000144 & 0.000144 \\
X$_{Mg^{24}}$ & 0.000496 & 0.000496 & 0.000496 & 0.000511 & 0.000510 & 0.000495
& 0.000510 & 0.000510 \\
X$_{Mg^{25}}$ & 0.000068 & 0.000068 & 0.000068 & 0.000070 & 0.000070 & 0.000068
& 0.000070 & 0.000070 \\
X$_{Mg^{26}}$ & 0.000081 & 0.000081 & 0.000081 & 0.000084 & 0.000084 & 0.000081
& 0.000084 & 0.000083 \\
Z/X & 0.026988 & 0.026988 & 0.025534 & 0.026217 & 0.027787 & 0.027077 &
0.026260 & 0.026296 \\
\noalign{\smallskip}
\hline
\end{tabular}
\end{flushleft}
\end{table*}

Element segregation represents in fact a competition between two kinds of
processes.
First the atoms move under the influence of external forces
(due to gravity, radiation, etc.), second they collide with other atoms
and share the acquired momentum with them in a random way, which slows
down their motion. This competition leads to element stratification and
decreases the entropy.

The computations of microscopic diffusion are based on the Boltzmann
equation for dilute collision-dominated plasmas. At equilibrium the
solution of the equation is the maxwellian distribution function: $f=f(0)$.
In stars the distribution is not maxwellian,
but the deviations from the maxwellian distribution are very
small. Two different methods are developped to solve the Boltzmann equation in
this framework. The first method relies on the
Chapman-Enskog procedure (described in Chapman and Cowling, 1970), with
convergent series of $f$ computed with successive approximations. This method
is used, for example, by Bahcall and Loeb (1990),
Proffitt and Michaud (1991), Michaud and
Vauclair (1991), Bahcall and Pinsonneault (1992) and 
CVZ, in which a complete description of
the numerical schemes may be found.

\begin{figure*}
\begin{center}
\epsfxsize=18.5cm
\epsfysize=18.5cm
\epsfbox{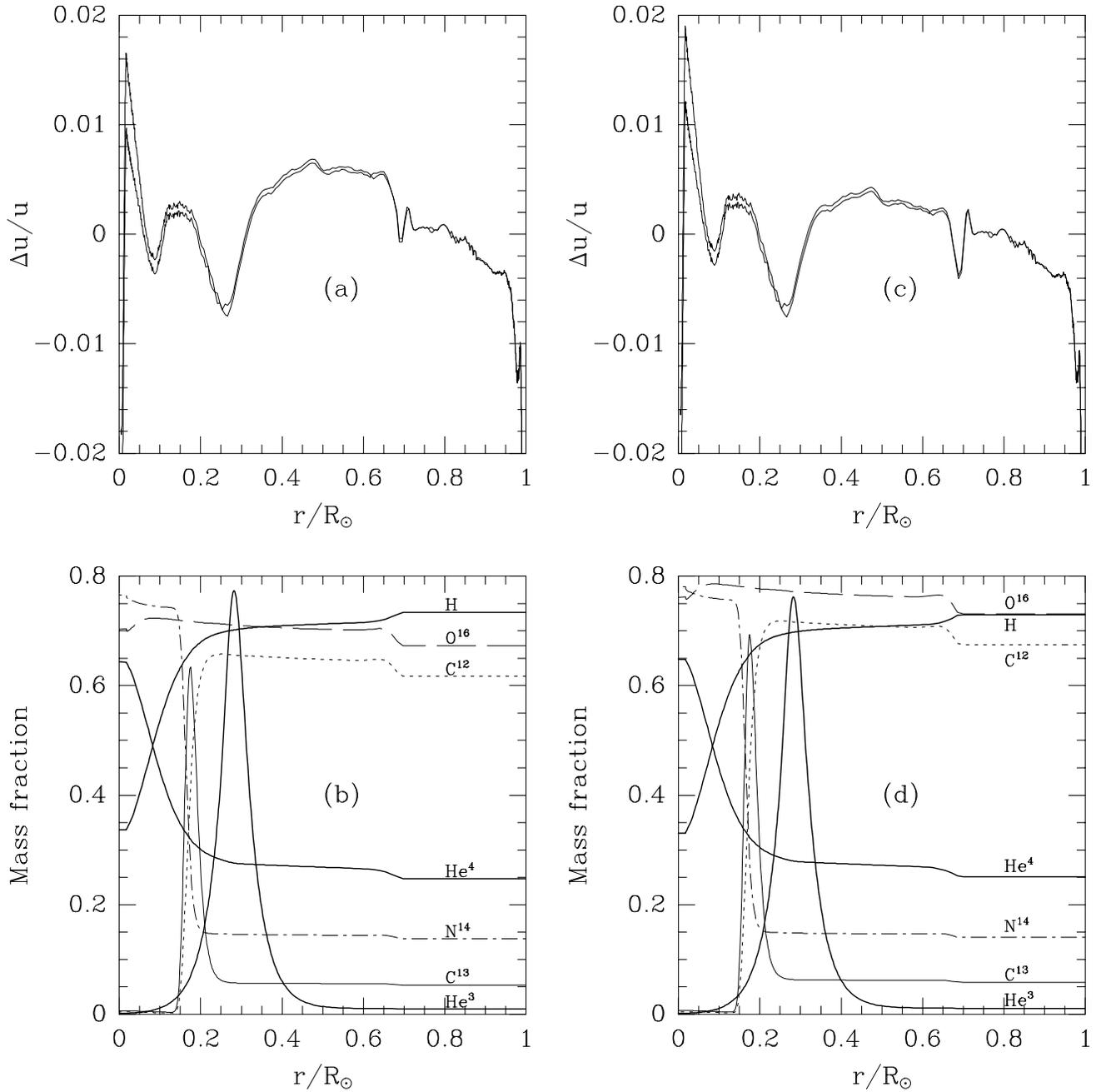}
\end{center}
\caption{
Models with pure element segregation. The presentation is 
the same as in figure~1. The left graphs represent the model 
obtained with initial abundances as given by Grevesse (1991) 
(model~2). The right graphs represent an interated model  
computed so that the final abundances are very close to 
Grevesse's observed ones (table~4). The $ \Delta u_{s} $ are 
clearly smaller for the iterated model.
The abundance variations are shown in the bottom graphs.
The relative abundance variations in the convective zone 
compared to the interior (depletion of all the elements 
except hydrogen) are the same in figures~2b and 2d, but the 
absolute values differ, so that in fig.~2d the outer values 
are close to those given in fig.~1b for all elements except 
helium 4, which is adjusted in the calibration of the model.
 } \end{figure*}

The second method is that of Burgers (1969), in which separate flow and
heat equations for each component of a multi-component mixture are
solved simultaneously.  Descriptions of this method may be found for
example in Cox,
Guzik and Kidman (1989), Proffitt and VandenBerg (1991), Thoul, Bahcall and
Loeb (1994), Richer and Michaud (1993). This method does not include for the
moment the problem of partial ionisation which has been studied within
the framework of the Chapman and Cowling method (Montmerle and Michaud
1976; Vauclair, Hardorp, Peterson, 1979; Alecian and Vauclair, 1981).

In the present paper the first method has been used for the treatment 
of the Boltzmann equation, as described in CVZ.
For collisions between charged ions,
the Paquette et al. (1986) method has been introduced. The basic question
concerns the divergence of the coulomb interaction cross
sections. In the first computations of diffusion, the ``Chapman and
Cowling approximation'' was used, assuming a cut-off of the cross section
equal to the Debye shielding length.  Paquette et al. (1986)
proposed a more precise treatment of this problem,
with a screened coulomb
potential in which the characteristic length is taken as the largest of the
Debye
length and interionic distance.The Paquette et al. tables of collision
integrals have been extensively used in the present paper.

The radiative acceleration on the elements  are
not included in these computations. From crude approximations,
we suppose that they are negligible in solar type
stars (Michaud et al., 1976), but this should be tested
in the future, as more precise computations may lead to larger values
(Michaud, 1987).

\begin{table*}
\caption{Predicted neutrino flux in 10$^{10}$ cm$^{-2}$.s$^{-1}$.}
\begin{flushleft}
\begin{tabular}{cccccccc}
\hline\noalign{\smallskip}
 \multicolumn{1}{c}{}
& \multicolumn{1}{c}{$\Phi$(pp)}
& \multicolumn{1}{c}{$\Phi$(pep)}
& \multicolumn{1}{c}{$\Phi$($^{7}$Be)}
& \multicolumn{1}{c}{$\Phi$($^{8}$B)}
& \multicolumn{1}{c}{$\Phi$($^{13}$N)}
& \multicolumn{1}{c}{$\Phi$($^{15}$O)}
& \multicolumn{1}{c}{$\Phi$($^{17}$F)}\\
\noalign{\smallskip}
\hline\noalign{\smallskip}
Model 1 & 6.00 & 1.39 $10^{-2}$ & 0.45 & 5.45 $10^{-4}$ & 4.52 $10^{-2}$
& 3.85 $10^{-2}$ & 4.90 $10^{-4}$ \\
Model 2 & 5.96 & 1.39 $10^{-2}$ & 0.47 & 5.98 $10^{-4}$ & 5.05 $10^{-2}$
& 4.31 $10^{-2}$ & 5.53 $10^{-4}$ \\
Model 3 & 5.94 & 1.38 $10^{-2}$ & 0.48 & 6.38 $10^{-4}$ & 5.79 $10^{-2}$
& 4.98 $10^{-2}$ & 6.40 $10^{-4}$ \\
Model 4 & 5.96 & 1.39 $10^{-2}$ & 0.47 & 6.06 $10^{-4}$ & 5.09 $10^{-2}$
& 4.35 $10^{-2}$ & 5.60 $10^{-4}$ \\
Model 5 & 5.94 & 1.38 $10^{-2}$ & 0.48 & 6.33 $10^{-4}$ & 5.59 $10^{-2}$
& 4.81 $10^{-2}$ & 6.18 $10^{-4}$ \\
\noalign{\smallskip}
\hline
\end{tabular}
\end{flushleft}
\caption{Predicted neutrino capture rates for the chlorine and gallium experiments in SNU.}
\begin{flushleft}
\begin{tabular}{lllllllllll}
\hline
\noalign{\smallskip}
 \multicolumn{1}{c}{Neutrino}
& \multicolumn{2}{c}{Model 1}
& \multicolumn{2}{c}{Model 2}
& \multicolumn{2}{c}{Model 3}
& \multicolumn{2}{c}{Model 4}
& \multicolumn{2}{c}{Model 5}\\
\noalign{\smallskip}
\cline {2-11}\noalign{\smallskip}
 source & $(\Phi \sigma)_{Cl}$ & $(\Phi \sigma)_{Ga}$ & $(\Phi
\sigma)_{Cl}$ & $(\Phi \sigma)_{Ga}$ & $(\Phi \sigma)_{Cl}$ & $(\Phi
\sigma)_{Ga}$ & $(\Phi \sigma)_{Cl}$ & $(\Phi
\sigma)_{Ga}$ & $(\Phi \sigma)_{Cl}$ & $(\Phi \sigma)_{Ga}$ \\
\hline
pp & 0 & 70.722 & 0 & 70.353 & 0 & 70.042 & 0 & 70.303 & 0 & 70.068 \\
pep & 0.223 & 2.992 & 0.223 & 2.992 & 0.221 & 2.965 & 0.223 & 2.993 &
0.221 & 2.973 \\
$^7$Be & 1.064 & 32.722 & 1.117 & 34.342 & 1.149 & 35.352 & 1.124 &
34.574 & 1.146 & 35.255 \\
$^8$B & 5.781 & 13.252 & 6.34 & 14.534 & 6.768 & 15.515 & 6.42 &
14.718 & 6.706 & 15.372 \\
$^{13}$N & 0.075 & 2.792 & 0.084 & 3.118 & 0.096 & 3.576 & 0.084 & 3.143
& 0.093 & 3.456 \\
$^{15}$O & 0.254 & 4.46 & 0.284 & 4.994 & 0.329 & 5.775 & 0.287 & 5.045 
& 0.318 & 5.581 \\
$^{17}$F & 0.003 & 0.057 & 0.004 & 0.065 & 0.004 & 0.075 & 0.004 & 0.066
& 0.004 & 0.072 \\
Total & 7.4 & 126.997 & 8.052 & 130.398 & 8.567 & 133.3 & 8.142 &
130.842 & 8.488 & 132.769 \\
\noalign{\smallskip}
\hline
\end{tabular}
\end{flushleft}
\end{table*}
\subsection{Rotation-induced mixing}

We introduced in our computations the rotation-induced mixing 
as prescribed by Zahn (1992). In a rotating star, due to 
centrifugal effects, the gravity equipotentials are no more 
spherical, which induces a circulation of matter between 
polar and equatorial regions: the so-called meridional 
circulation. This circulation itself induces a transport of 
angular momentum, thereby creating shears which become 
unstable in the horizontal direction, while the vertical 
shears are stabilized by the density gradient. This large 
scale horizontal turbulence decays into small scales and 
becomes 3D when the turnover rate of the turbulence exceeds 
the angular velocity.

Meanwhile the horizontal turbulence ``cuts down'' the effect 
of advection on the transport of the chemical species, as 
the elements which go up in the upward flow of matter can be 
transported into the downward flow by horizontal motions 
before reaching the top layers. The transport of angular 
momentum is more efficient than the transport of chemicals. 
In the limit of extremely large horizontal diffusivity, the chemical 
composition is constant along a level surface, and the 
transport of chemicals is negligible. The angular momentum 
behaves differently as, when the rotation velocity is 
constant along a level surface, the angular momentum is not, 
so that it is still transported.

The horizontal transport of angular momentum smoothes out 
the original meridional circulation. Taking this feedback 
effect into account, Zahn (1992) showed that the whole 
process is stopped within an Eddington-Sweet time-scale, 
unless angular momentum is extracted from the star due to a 
wind.

In case of a moderate wind which extracts angular momentum 
at the rate $ \left( {dJ \over dt }\right) $, an asymptotic 
regime is reached with a circulation velocity as a function 
of radius given by (following Zahn (1992) eq.4.15):
  $$  
U  (r) =
 {5 \over \rho {(r)} \; r^{4}  \; \Omega } \;
 {3 \over 8\pi  } \>
\left( {dJ \over dt }\right)
\eqno(1)
  $$
where $ \rho  $ is the local density, or:
  $$  
U(r) =
 {5 \over 2 } \;
 {1 \over \Omega (r) \; r \; M(r) } \;
 {\rho _{m} \over \rho (r) } \;
\left(   {dJ \over dt }\right)
\eqno(2)
  $$
where $ \Omega (r) $ is the local angular velocity, $ M(r) $ the 
mass inside radius $ r $ and $ \rho _{m} $ the average density 
inside radius $ r $.

The effective diffusion coefficient is then expressed by 
Zahn's eq.~4.21 :
  $$  
D_{eff} =  {C_{h} \over 50 } \;  {r \vert U(r)\vert \over  \alpha } \eqno(3)
  $$
where $ C_{h} $ is a parameter related to the horizontal 
viscosity $ (C_{h} \la 1)$  and $ \alpha  $ is related to the 
differential rotation:
  $$  
\alpha  =  {1 \over  2} \;  {d \ln \, (r^{2}\Omega ) \over  d \ln \, r } \eqno(4)
  $$

If the deviation of $ \Omega  $ from solid rotation is neglected
$ (\alpha  = 1) $,
and if $ \Omega  $ is supposed to decrease with time following a
``Skumanich law'' for which $ \Omega  \propto t^{- {1 \over 2 }} $ we 
obtain:
  $$  
 {dJ \over  dt}  \propto \Omega ^{3} \eqno(5)
  $$
and the  effective 
mixing coefficient is of the form:
  $$  
D_{eff} \simeq r. U(r)  \propto  {\Omega ^{2} \over \rho  \; r^{3} }
\eqno(6)
  $$
Equation~(3) has been used in the present computations
with $ \alpha  = 1 $. The 
proportionality factor 
$ C_{h} $
has been adjusted to obtain the right 
lithium depletion in the Sun (section~4).

\subsection{Stabilizing $ \mu  $-gradients}

Mixing processes in stars may be stabilized in the regions 
where the mean molecular weight rapidly decreases with 
increasing radius. This occurs specially in the nuclear 
bruning core: we can thus infer that the rotation-induced 
mixing becomes inefficient as soon as the $ \mu  $-gradient 
becomes larger than some critical value. This question has 
been discussed by several authors (Mestel (1965),  
Huppert and Spiegel (1977)). Although no precise value can be given 
for this critical $ \mu  $-gradient, an order of magnitude can 
be obtained from simple considerations.

Huppert and Spiegel (1977) suggest that mixing can 
penetrate the nuclear burning core within a scale height 
given by:
  $$  
h \simeq r \;  {\Omega (r) \over N_{\mu } } \eqno(7)
  $$
where $ r $ is the local radius, $ \Omega (r) $ the angular 
rotation velocity and $ N_{\mu } $ the buoyancy frequency due 
to the $ \mu  $-gradient.
  $$  
N_{\mu }^{2} \simeq
 {GM(r) \over r^{2} } \;
\left\vert  {d\ln \, \mu  \over  dr }\right\vert \eqno(8)
  $$

 From eq.~(7) we can derive a critical $ \mu  $-gradient 
obtained by specifying that  $ h $ must be a small
fraction of $ r $ ($ h = \varepsilon  \ r $):
  $$  
(\nabla \ln \, \mu )_{c} \simeq
 {1 \over \varepsilon ^{2} } \, {r^{2}_{c} \; \Omega ^{2} \; (r_{c})\over  GM (r_{c})} \eqno(9)
  $$
where all the quantities should be computed at the place 
where the actual $ \nabla \ln \, \mu  $ is equal to the 
critical one.

With values of $ r_{c} $ between 0.1~$R_{\odot}$ and 0.2~$R  
_{\odot}$ and values of $ \Omega  $ between $ 3 \times  10^{-6} $ 
$ (V_{R_{\odot}} \simeq 2 $~km.s$ ^{-1}) $ and $ 10^{-4} $
$ (V_{R_{\odot}} \simeq 70 $~km.s$ ^{-1}) $ we find, for
$ \varepsilon  = 0.1 $:
  $$  
4 \times  10^{-15} < (\nabla \ln \, \mu )_{c}  < 4 \times  10^{-12}
  $$
Solar structure 
computations lead to stronger constraints on the critical $  
\mu $-gradient. With $ (\nabla \ln \, \mu )_{c} \la 10^{-14} $ no 
mixing could occur in the Sun after 0.1~Gyr. With 
$( \nabla \ln \, \mu )_{c} \ga 10^{-12} $ too much mixing would 
occur in the core, and the consistency with helioseismology 
would be lost (Gaig\'e 1994). Our best model is obtained with $ (\nabla \ln 
\, \mu )_{c} = 4 \times  10^{-13}$ (section~4.3).

\begin{figure*}
\begin{center}
\epsfxsize=18.5cm
\epsfysize=18.5cm
\epsfbox{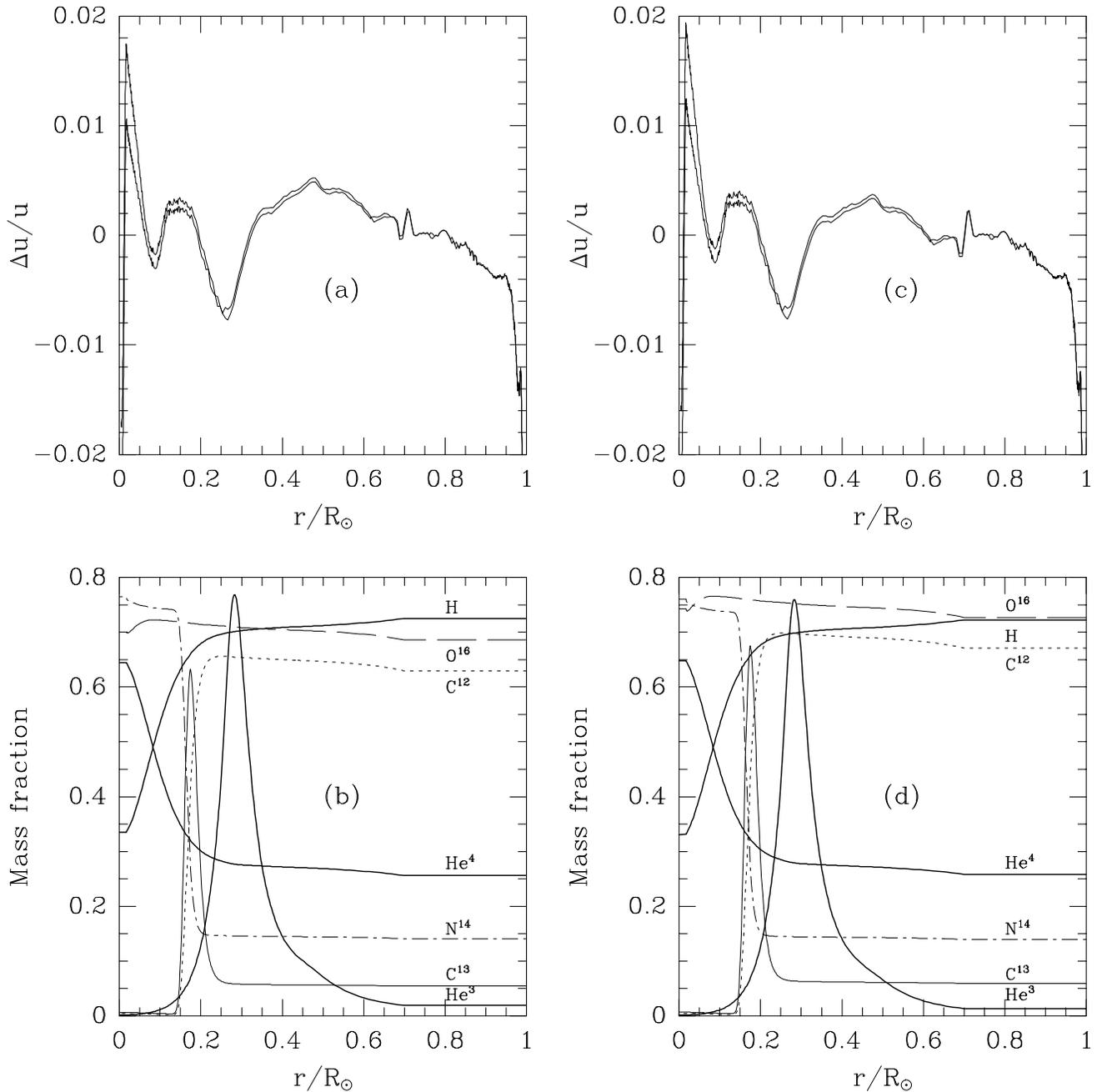}
\end{center}
\caption{
Models with element segregation and rotation-induced
mixing. The presentation is the same as in figures~1 and 2.
The left graphs correspond to models computed with initial
Grevesse (1991) abundances (model~4).
The right graphs correspond to
iterated models so that the final abundances are
close to the observed
Grevesse (1991) abundances
(model~5).
Here the parametrisation of the rotation-induced mixing is
that proposed by Zahn (1992) in the asymptotic regime. The
mixing efficiency is supposed annihilated
when the $ \mu  $-gradient is larger than $ 4 \times  10^{-13} $~cm$ ^{-1} $ (
see
text). This cut-off is very important as a deeper mixing
inside the nuclear burning core destroys the consistency of
the computed and observed $ u $ values.
Here the $ \Delta u $ values are slightly smaller in figures~3a
and 3c than in figures~2a and 2c.
Meanwhile lithium and beryllium are destroyed as observed in
the Sun (see figure~4).}\end{figure*}

\subsection{The diffusion routine}

The stellar evolution code used in these computations is
the Geneva code, described several times in the literature. 
The system of nuclear reactions and the abundance variations 
in the standard models are computed as in Maeder (1983). 
They are separatly determined for 15 isotopes : H, $ ^{3} 
$He, $ ^{4} $He, $ ^{12} $C, $ ^{13} $C, $ ^{14} $N, $ ^{15} $N, 
$ ^{16} $O, 
$ ^{17} $O, 
$ ^{18} $O, 
$ ^{20} $Ne, 
$ ^{22} $Ne, 
$ ^{24} $Mg, 
$ ^{25} $Mg, 
$ ^{26} $Mg.
The heavier elements are combined in a single mass fraction 
Z.

We have added in this code a diffusion routine for each 
isotope in a similar way as described in CVZ. The diffusion 
equations are written in lagragian coordinates as :
  $$  
 {\partial c_{i} \over \partial t } =
D_{i} \;
 {\partial ^{2}  c_{i} \over \partial  \, m^{2} }
+ E_{i}  \; 
 {\partial  \, c_{i} \over \partial  \, m^{2} } +
F_{i} \; c_{i} \eqno( 10)
  $$ 
where $ c_{i} $ stands for the concentration of isotope $ i 
$, and:
  $$  
\matrix {
& D_{i}  & =  (4\pi  r^{2}\rho )^{2}  \; (D_{eff} + D^{i}_{s}) & \cr
& E_{i}  & = (4\pi  r^{2}\rho )^{2}  \; \left(4\pi  \,  {\partial r^{2}\rho  \, 
(D_{eff} \; D^{i}_{s}) \over  \partial m } - V_{i}\right) & (11)  \cr
& F_{i}  & = - \lambda _{i} - 4 \pi  \;  {\partial  \, (r^{2} \rho  V_{i})\over \partial m } &  \cr
          }
  $$

Here $ D_{eff} $ is the effective mixing coefficient (the 
same for all the isotopes) while $ D^{i}_{s} $ is the 
segregation coefficient, computed for each isotope using 
Paquette et al. (1986) tables. $ V_{i} $ represents the 
segregation (microscopic) velocity. The nuclear destruction 
rate $ \lambda _{i} $ is only included in $ F_{i} $ for lithium
and 
beryllium, which are treated separately from the 
network.

For the 15 isotopes included in the nuclear network, the 
computation procedure is the following:

-- at each evolutionary step, 
 equation (10) is solved separately for all the 
isotopes except hydrogen. The method is the same as 
described in CVZ: a Cranck-Nicholson scheme with the 
inversion of a tridiagonal matrix including all the mesh 
points down to the center

-- this diffusion routine is used with a smaller time step 
as the evolution time step, for a better precision. 
Typically 20 resolutions of the diffusion equation are done 
between two computations of a complete model

-- the new mass fractions of each isotope are computed, 
taking into account the normalisation equation:
  $$  
X_{1} +
\sum _{k \not= 1} \;
X_{k} + Z = 1
  $$
Satisfying this equation needs a consistent resolution of the 
abundances of all the considered isotopes with the inversion 
of a $ (k + 1) $ order matrix

-- the abundance variations due 
to the nuclear reactions are then computed.

\section{Helioseismological constraints}
We shall compare the $ u \ (r) =  {P \ (r) \over  \rho  \ (r)} $ 
function and the $ Y_{surf} $ 
quantity in our theoretical models with those in the seismic 
model derived from the observed p-mode frequencies with the method 
described by Dziembowski et al. (1994). The seismic model 
that we adopt here differs from the one presented in that 
paper in two respects. First we use new frequency data for $  
 \ell \le 3$  degrees from BISON network (Elsworth et al., 
 1994). This has a most noticeable effect in $ u \ (r) $ in the 
 core. The second is the use of OPAL rather than MHD 
 equation of state (EOS) in the reference model adopted for the 
 inversion (for consistency with the opacities),
 which does not change the $ u $ values in a 
 significant way. The only important consequence of this 
 change is a  somewhat higher seismic value for $ Y_{surf} 
 $.

Within the adiabatic part of the convective zone the sound 
speed is determined solely by the value of the $ M $ to $ R 
$ ratio and by the value of adiabatic exponent, $ \Gamma_{1} 
$. Comparison of the model and the seismic $ u \ (r) $ in this 
region provides therefore a test of the EOS. The situation 
in the radiative interior is more complicated. Let us note 
that approximately $ u \propto  {T \over \mu } $, and therefore 
the radiative transport of energy and the element diffusion 
directly affect $ u \ (r) $  below the convective envelope.

\begin{figure} 
\begin{center}
\epsfxsize=9.5cm
\epsfysize=18.5cm
\epsfbox{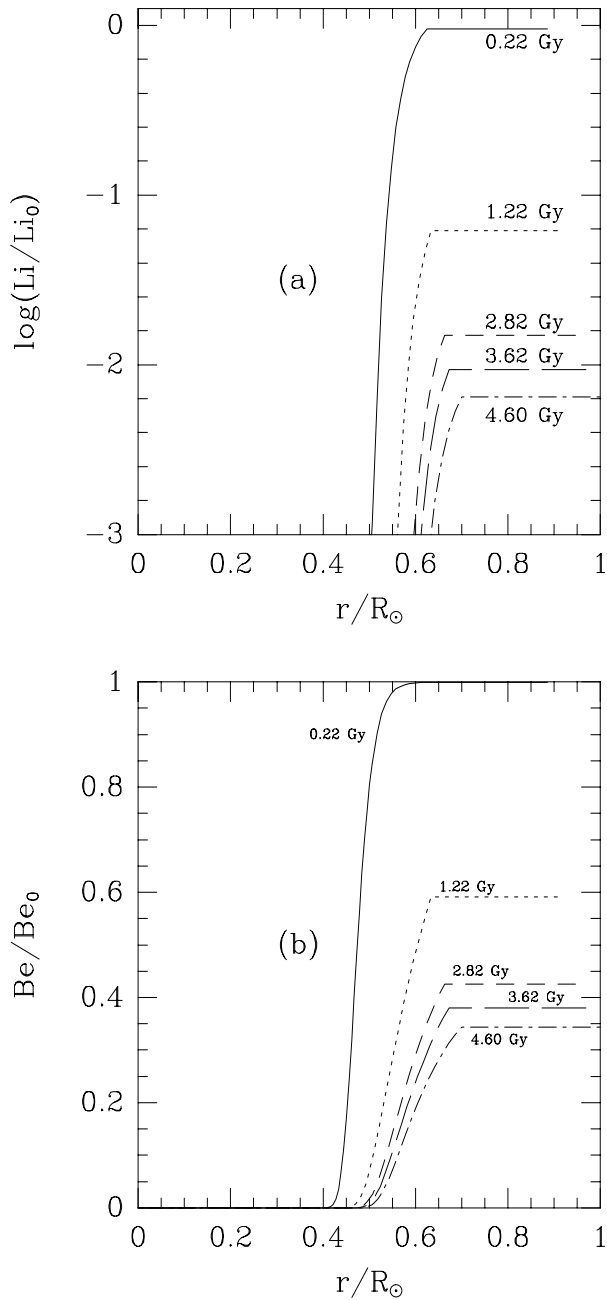}
\end{center}
\caption
{Lithium (top) and beryllium (bottom) variations profiles in 
the best model~5 as a function of age. At the age of the Sun 
lithium is destroyed by 155 and beryllium by 2.9.
 }
\end{figure}

In general, it is not possible to disentangle the 
uncertainty in the opacity and in the diffusion 
coefficients. Only the convective overshooting leading to a 
discontinuity in $  {du \over  dr} $ leaves a signature 
in the sound speed which 
is seismically detectable. Monteiro et al. (1994) looked for 
it and came up  with an upper limit on the extent of 
such overshooting, which is 0.07 of the local pressure 
distance scale or $ 0.006  $~R$ _{\odot} $. In our models such 
overshooting has been ignored and the above results show 
that this is a good approximation.
The bottom of the convective zone in such case may be 
determined quite accurately. In the seismic model used in 
this work $ r_{cz} = (0.7137 \pm 0.0002)$~R$_{\odot}$, which 
agrees very well with the first accurate helioseismic value
$ r_{cz} = (0.713 \pm 0.003)  $~R$_{\odot}$
(Christensen-Dalsgaard et al.,1993).

There are still some uncertainties due to the opacities.
Unfortunately, we do not have a good way to assess limits for 
the induced modifications. Alternative opacities (OP, Seaton 
et al. 1994) are by about 40 percent lower at temperature 
and density ranges of interest here. However, this large 
difference may perhaps be due to a neglect of the plasma effects 
on the atomic properties, in the OP calculations, and therefore 
may not serve as an estimate of the uncertainty.

In such a situation it is important to make use of the 
additional constraint which is the value of $ Y_{surf} $ 
determined by means of helioseismology. In this comparison 
the uncertainty in the opacity data is rather unimportant. 
The possibility of a seismic measure of helium abundance rests 
on a large value of the $ \Gamma_{1} $ derivative with 
respect to $ Y $ in the HeII ionization zone and 
obviously requires very accurate thermodynamical data. The 
seismic values obtained with Dziembowski et al. (1994) 
method of inversion for two versions of the EOS adopted in 
the reference model are

$ Y_{surf}  = 0.2440 \pm 0.0003$ for MHD

$ Y_{surf}  = 0.2505 \pm 0.0003$ for OPAL

The test of $ \Gamma_{1} $ in the lower convective zone 
points to OPAL data as more accurate, which should be 
expected as the OPAL EOS is obtained in a more fundamental 
way. We stress that the model values of $ Y_{surf} $ is 
quite insensitive to the choice of the EOS.
The errors given above reflect only the frequency errors 
quoted by  the observers.
The actual uncertainty in the seismic values is much larger. 
For instance, Basu and Antia (1995) find 0.246 and 0.249 for 
the corresponding quantities. The results are indeed 
depended on the adopted method of inversion. In 
particular they depend on  smoothing the $ \Delta u \over u $ 
function which describes the relative differences between 
the solar and the model $ u \ (r) $.
Without smoothing, which is perhaps a better choice is the 
only goal is the determination of the He abundance, our 
method yields $ Y_{surf} = 0.2548$ for the OPAL EOS. 
(Pamyatnykh, private communication). The problem certainly 
requires further examination because, as we shall see, it is 
essential to reduce the uncertainty to the $ 10^{-3} $ level.

\section{The results}
Five solar models have been computed and compared to the helioseismological
sun (Figures~1 to 3). The values of the characteristic parameters of
the models, abundances and neutrino production are given in 
tables 2 to 6.

\subsection{The ``standard'' model}

This model includes the physics as 
discussed in section~2-1, with no element
segregation and no other mixing than inside the convection zone. 
The comparison
with helioseismology is presented
 in figure~1a, and the fractional abundance of
several interesting elements in figure 1b. The two curves in figure 1a
correspond to the uncertainty in the helioseismological 
inversion. 

A comparison of seismic values of $ r_{cz} $ and $ Y_{surf} 
$ with those for Model~1, given in Tables~2 and 3 reveals 
large differences.
Also large differences are seen in the sound speed behavior 
throughout the whole interior as seen in Fig.~1. We do not 
pay attention to the differences in the outer part of the 
convective zone. They may be accounted for by inadequacies 
of the mixing length theory (outermost part) and in the MHD 
equation of state. Our main concern is the value $ \Delta u 
\over u $ of about 0.015 in the outer part of the radiative 
interior, which is too large.

\begin{figure*}
\begin{center}
\epsfxsize=18.5cm
\epsfysize=18.5cm
\epsfbox{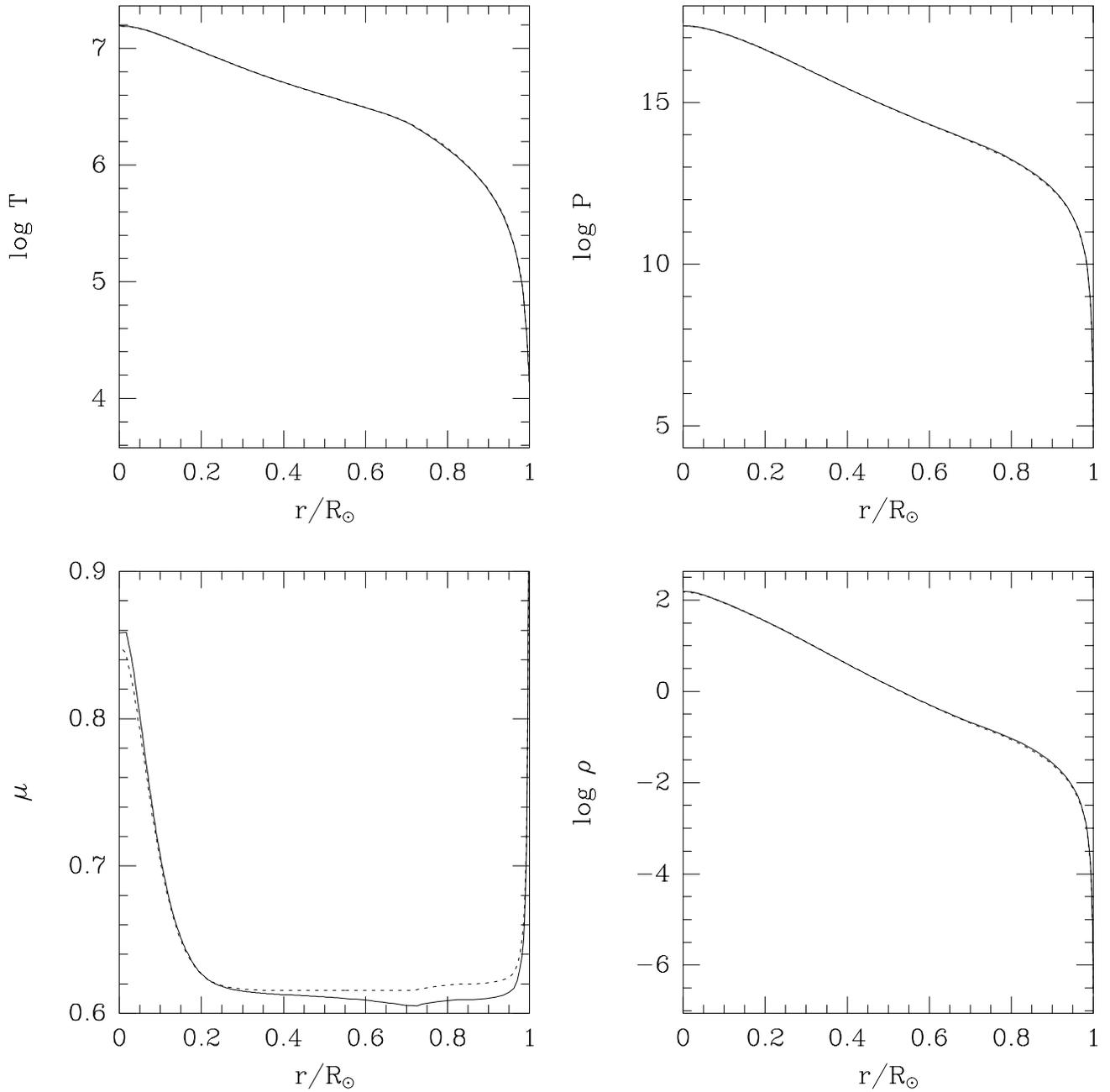}
\end{center}
\caption
{Comparison between the parameters log T, log P, $ \mu  $ and 
log~$ \rho  $ in our ``best model''~5 (solid lines) and the 
standard model (dashed lines). While no clear difference can 
be seen at these scales for T, P and $ \rho  $, the difference 
in $ \mu  $ due to the element segregation is clearly visible. 
It is the basic reason for the better agreement of this 
model with the helioseismological sound speed, compared to 
the standard model.}\end{figure*}

Of course
lithium is not depleted in model 1, neither is beryllium, and the solar
neutrino fluxes are too high compared with the results of the solar
neutrino detectors (tables~5 and 6).

\subsection{Models with pure element segregation (no mixing)}

-- the first of these models (model 2) has been obtained with element
segregation of helium and the initial abundances as given by Grevesse
(1991).

We see at once, from figure 2a, that the discrepancy below the convective
zone is considerably reduced compared to model~1, 
while figure 2b shows the influence of
segregation on the fractional abundances of the elements. The basic reason
for the improvement in the 
$u$ value is due to helium diffusion, which leads
to a diminution of the helium abundance of about 
10\% inside the convection
zone and below, thereby decreasing the local mean molecular weight.

-- model 2 is however inconsistent with the observed present element
abundances due to the segregation. Model 3 is similar 
to model 2, but
iterated so that the final abundances correspond to the values given by
Grevesse (1991). Table~4 shows the final abundances obtained, compared to the
observations.
 It is also interesting to compare figure 2d
with figure 2b to see the difference in the fractional abundances of the
elements.

There is a dramatic improvement in the agreement with the 
seismic model once the element segregation is introduced. In 
Model~3 the coincidence of $ r_{cz} $ and $ Y_{surf} $ with 
the seismic values is so perfect that, admittedly, it may be 
to some extent coincidental.

From tables 2 to 6 we check that models 2 and 3 are well calibrated, while
the neutrino fluxes are not decreased, as expected.

We insist on the fact that these models are computed without any arbitrary
parameters: element segregation is a simple consequence of the stellar
physics, with no special assumption added. They should not be considered as
non-standard models, but as improved standard solar models.

Lithium and beryllium are not more depleted than helium in these models, as
the depletion is only due to element segregation. 
It is necessary to
introduce a mild mixing below the convection zone to account for the
observed abundances of these elements.

\subsection{Models with element segregation and rotation-induced 
mixing}

Rotation-induced mixing as described in section 2 have been added in the
computations of models 4 and 5. Here a parameter has been ajusted (namely
the $C_{h}$ factor in Zahn's prescription) so as to obtain a lithium depletion
as observed. In consequence these two models must be considered as
non-standard.

In both models the mixing coefficient has been cut off at the core for 
a $ \mu  $-gradient of $ 4 \times  10^{-13} $. 
Models with mixing down to the solar core lead to large
discrepancies with helioseismology.
On the other hand, in our models, the mixing is inefficient in
the region of energy production in the Sun, and the 
consistency with helioseismology is even better than for 
models without mixing (see  figures~3a and 3c).

Model 4 is obtained with initial abundances as in Grevesse (1991), while
model 5 has been iterated so that the final abundances are those of
Grevesse (1991) (see table 4). It is interesting to note that model 5 is
still closer to 
the helioseismological Sun that model 4: in all cases improving the
physics leads to better results compared to the observations, which is very
encouraging.

While an inclusion of the element mixing somewhat improves 
the agreement in the $ u \ (r) $, it slightly destroys  the 
perfect coincidence in $ r_{cz} $ and $ Y_{surf} $. We do 
not consider any of these differences as really significant. 
Let us note that the seismic value of $ Y_{surf} $ obtained 
without regularization, which we regard as more credible, is 
right in the middle of the values for model~3 and 5. The 
difference between the two models is 0.005. An accuracy of 
0.001 in seismic $ Y_{surf} $ would thus yield an 
interesting constraint on the element mixing. The data 
certainly allow such accuracy.
The problems lie in data analysis and in reliability of the 
EOS.

In model 5 lithium is depleted with a ratio 1/155 and 
beryllium with a ratio 1/2.9, which is very close to the 
observations. The lithium and beryllium abundance profiles 
for various times are given in figure~4.

The difference in $ \mu  $-values from models~1 and 5 is given 
in figure~5.

\section{Conclusion}
Model 5 is presently our best model. 
It is a well calibrated model, including element segregation and a
parametrized rotation-induced mixing. 
It leads to a very good fit between
the computed 
$ u =  {P \over \rho  } $
function and that deduced from helioseismology below
the solar convection zone. The consistency is also very good inside the
convection zone except at the surface where a better mixing treatment
should be introduced. Lithium and beryllium are depleted as 
observed. The neutrino fluxes remain too high.

Below a radius of 0.4 of the solar radius a discrepancy remains
in the $ u $ curves. Although
the uncertainty on $u$ is much larger in the core than at the surface, these
features seem significant. This specific problem will be addressed in a
forthcoming paper.

Element segregation certainly takes place in the Sun's 
interior. It involves no nonstandard physics and its 
occurence is fully confirmed by the results of helioseismic 
inversion. On the other hand the macroscopic mixing is a 
hypothetical effect which provides a natural explanation of 
the Li and Be deficit but demands convincing observational 
confirmations and constraints. At present stage we may only 
conclude that there is no conflict between models 
reproducing solar Li and Be abundances and helioseismic 
data. There are good prospects for obtaining stringent 
constraints on the hypothetical mixing processes from 
helioseismology. On the road to this goal we regard as most 
important to improve reliability and accuracy of the He 
abundance determination. Progress should also be made in 
assessing uncertainties in the opacity data, so that the 
information about the sound speed behavior 
may lead to a more direct probe of the He 
distribution in the outer part of the radiative interior.

\bigskip

\noindent
{\it Note\/}:  Model~5 is available  on
request by electronic mail at the address:
richard@obs-mip.fr.

\begin{acknowledgements}We thank Alosha Pamyatnykh for his collaboration on 
the new version of the seismic model and for a helpful 
discussion. The work was supported, in part, by 
PICS ``France-Pologne''.
\end{acknowledgements}

\end{document}